\documentclass[aps,prl,twocolumn,showpacs,amsmath,preprintnumbers,amssymb,reprint]{revtex4-1}
\usepackage{graphicx}
\usepackage{dcolumn}
\usepackage[latin1]{inputenc}
\usepackage{color}
\usepackage{ulem}
\usepackage{amssymb}
\begin{document}
%
\title{Spectroscopy and Dynamics of a Two-Dimensional Electron Gas
on top of Ultrathin Helium Films on Cu(111)}
\author{N.~Armbrust}
\author{J.~G{\"u}dde}
\author{U.~H{\"o}fer}
\affiliation{Fachbereich Physik und Zentrum f{\"u}r
Materialwissenschaften, Philipps-Universit{\"a}t, 35032 Marburg,
Germany}
\author{S.~Kossler}
\author{P.~Feulner}
\affiliation{Physikdepartment E20, Technische Universit{\"a}t
M{\"u}nchen, 85747 Garching, Germany}

\date{\today}


\begin{abstract}                                                   %
%
Electrons in image-potential states on the surface of bulk
helium represent a unique model system of a two-dimensional
electron gas.
 Here, we investigate their properties in the extreme case of
reduced film thickness: a monolayer of helium physisorbed on a
single-crystalline (111)-oriented Cu surface.
 For this purpose we have utilized a customized setup for
time-resolved two-photon photoemission (2PPE) at very low
temperatures under ultra-high vacuum conditions.
 We demonstrate that the highly polarizable metal substrate
increases the binding energy of the first ($n=1$)
image-potential state by more than two orders of magnitude as
compared to the surface of liquid helium.
 An electron in this state is still strongly decoupled from the
metal surface due to the large negative electron affinity of
helium and we find that even one monolayer of helium increases
its lifetime by one order of magnitude compared to the bare
Cu(111) surface.
\end{abstract}
\maketitle

%
 Two-dimensional (2D) electron systems attract interest since more than
$40$~years.
 Apart from sheet structures such as graphene,
2D electron systems exist at heterostructures like
semiconductor-semiconductor \cite{Stormer79ssc,Davies98},
semiconductor-insulator \cite{Ando82rmp,Davies98,Monarkha04},
oxide-oxide \cite{Stemmer14annurev}, or metal-insulator
interfaces \cite{Gudde05pss}, and on the surface of condensed
materials with negative electron affinity
\cite{Cole73rmp,Andrei97,Monarkha04}.
 Two limiting cases are known:
dense electron layers with Fermi temperatures in the hundred K
range in quantum wells, particularly of semiconductor
heterostructures \cite{Stormer79ssc,Davies98}; and very dilute
2D electron gases in the image-potential states on top of
condensed Helium \cite{Cole73rmp,Andrei97,Monarkha04}
with Fermi temperatures in the mK range, and spacing of the
excited states in the microwave regime \cite{Grimes76prb}.
 For electrons on the bulk surface of He, the maximum density is small: for more than
$\approx 2\times 10^{9}$ electrons per cm$^{-2}$ the layer
becomes unstable \cite{Edelman80pu} and the electron gas
remains in the classical regime.
 Early, it was realized that the density of such electron
layers can be significantly increased by growing He films of
finite thickness on top of a substrate with large permittivity
\cite{Ikezi81prb,Peeters83prl}.
 For a $100$~\AA\ He film on a doped silicon substrate, for example,
densities of up to $10^{11}$ cm$^{-2}$ have been reported
\cite{Etz84prl}.
 Such densities offer the possibility to study the quantum
regime of this almost ideal 2D electron system including
effects as Wigner crystallization and quantum-melting
\cite{Peeters83prl} as long as the electron gas is well
decoupled from the substrate.
 For very thin films, however, the coupling to the substrate will
be strongly influenced by surface roughness and impurities of
the substrate which can lead to lateral localization and
enhanced tunneling through the film
\cite{Etz84prl,Gunzler96ss}.
 The study of this regime thus requires the combination of
advanced surface science and cryogenic techniques.

 In this Letter, we investigate the limiting case of a monolayer
(ML) of He on an atomically flat single-crystalline metal
substrate and present a study of the electron transfer dynamics
of the image-potential states on this archetypical 2D system.
 Image potential states on clean and rare gas covered metal
surfaces already proved to be ideal model systems for the
electron transfer dynamics at surfaces and through thin
dielectric layers, by theory and by experiment, in particular
by two-photon photoemission (2PPE) studies \cite{Echeni04ssr}.
 For thin films of the heavier rare gases, it
has been shown that the coupling of the image-potential states
to the metal strongly depends on the electron affinity of the
film which represents a tunnel barrier in the case of negative
electron affinity
\cite{Berthold02cpl,Marinica02prl,Berthold04apa,Gudde05pss}.
 He films are expected to exhibit a particular high barrier (the
electron affinity of condensed He is $-1.3$~eV
\cite{Sommer64prl}), which offers the possibility to create a
2D electron gas with high binding energy and very long lifetime
even for very small thicknesses.
 He films are also unique with regard to their structure.
 Whereas all other rare gases, including Neon, form islands at sub-monolayer coverage and
solid single-crystals at larger thicknesses
\cite{Schlic92ss,Wolf96prb,Berthold04apa,Berthold04ss},
sub-monolayer of He grow as uniform 2D gas layer and thick He
films are liquid because of the large zero-point energy and the
weak van der Waals interaction \cite{Note1}.
 Despite of these outstanding quantum properties,
the 2D electron gas on ultrathin He films was not investigated
until now, mainly because of experimental challenges.

 2PPE of He films requires very low temperatures
\cite{Nieder02prl,Nieder05prb}, shielding from thermal
radiation of the environment which leads to desorption of the
layers \cite{Nieder02prl,Nieder05prb}, and highly sensitive
electron spectroscopy at low laser intensities.
 Tackling these challenges with customized equipment, we present
results on binding energies and lifetimes of the image
potential states on ultrathin He films physisorbed on a
single-crystalline Cu(111) substrate.

\begin{figure}[t]
    \begin{center}
    \includegraphics[width = 0.9\columnwidth]{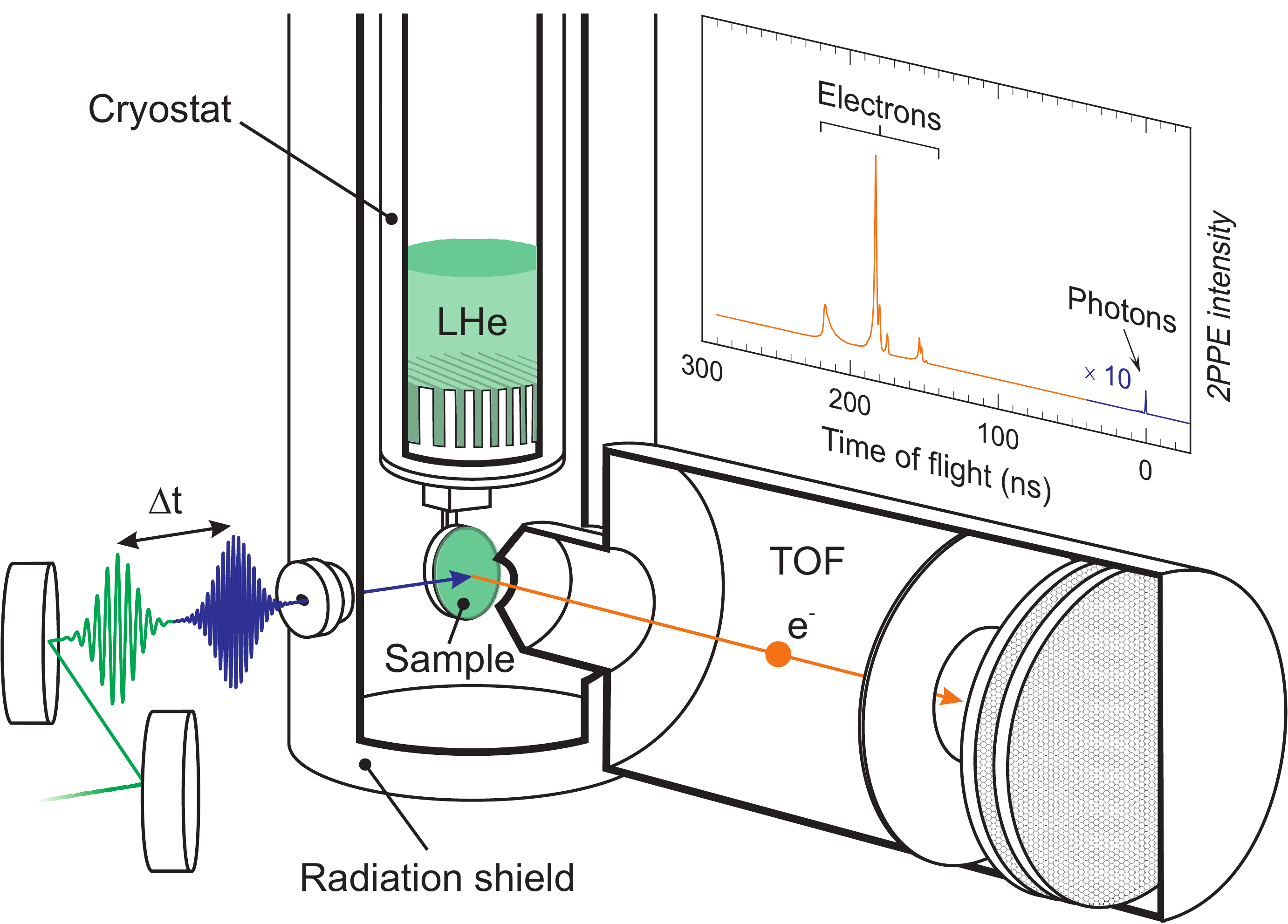}
    \caption[]{Cut-away view onto the cryostat, the
    sample mount, the TOF spectrometer and the light path. The 80~K
    radiation shields surrounding cryostat, sample mount and
    spectrometer were brought in seamless contact upon He layer
    preparation and data acquisition. The laser beams were focused
    onto the sample through a 2~mm$^2$ aperture in the radiation shield at
    80$^\circ$ angle of incidence. The
    inset shows a typical TOF spectrum with the consecutive arrival of
    scattered uv photons and photoelectrons.}
    \label{fig:fig:exp}
    \end{center}
\end{figure}

%
 Figure~\ref{fig:fig:exp} shows the experimental setup.
The liquid He (LHe) bath cryostat of our UHV chamber (base
pressure of $5\times10^{-11}$~mbar) was operated at a He
pressure of typically $5\times10^{-2}$~mbar corresponding to a
LHe temperature below 1~K \cite{Clement55pr}. 
 Well-defined Cu(111) surfaces have been prepared by {\it in-situ}
epitaxially growth of approximate hundred ML thick Cu films on
a Ru(0001) single crystal.
 The latter was thermally coupled to the cryostat by a
thin, spot-welded single-crystalline tungsten rod.
 With this setup, sample temperatures below 1.2~K were
obtained \cite{Kossler11dr,Kossler14prb}.
 He films were prepared by dosing purified He gas through a
capillary aiming at the sample.
 We compensated the unavoidable loss of He due to laser
stimulated desorption by a controlled continuous flow of He,
which allows for the variation of the adlayer density within
the laser spot.
 Further details of the sample preparation and characterization
are described in the Supplemental Material \cite{supplemental}.

 In order to avoid electronically stimulated desorption of the He
layers by the $300$~K black body radiation of the environment
\cite{Nieder02prl,Nieder05prb}, the whole cryostat and the
time-of-flight (TOF) spectrometer used for electron
spectroscopy ($\leq 20$~meV energy resolution) were surrounded
by radiation shields cooled to $80$~K by liquid nitrogen
(LN$_{2}$) \cite{Kossler11dr,Kossler14prb}.
 During the experiment, these shields
were completely closed except of a tiny 2~mm$^2$ aperture for
the laser beams [Fig.~\ref{fig:fig:exp}].

The laser setup provided p-polarized femtosecond laser pulses
in the visible (vis) and ultraviolet (uv) spectral range with
variable time delay $\Delta t$ at a repetition rate of
$250$~kHz \cite{Damm09prb}.
 Typical photon energies and pulse lengths of $\hbar\omega_{\rm vis} = 2.28$~eV,
$\tau_{\rm vis} = 50$~fs and $\hbar\omega_{\rm uv} = 4.56$~eV,
$\tau_{\rm uv} = 80$~fs enabled excitation of the first two
image-potential states with the uv pulses but simultaneously
avoided an excessive background signal due to one-photon
photo\-emission.
 The average laser power of each laser beam
was reduced to only $\sim 50$~$\mu$W ($\sim 0.2$~nJ/pulse) as a
compromise between He desorption and 2PPE intensity.

%
\begin{figure*}[t]
    \begin{center}
    \includegraphics[width = 0.9817\textwidth]{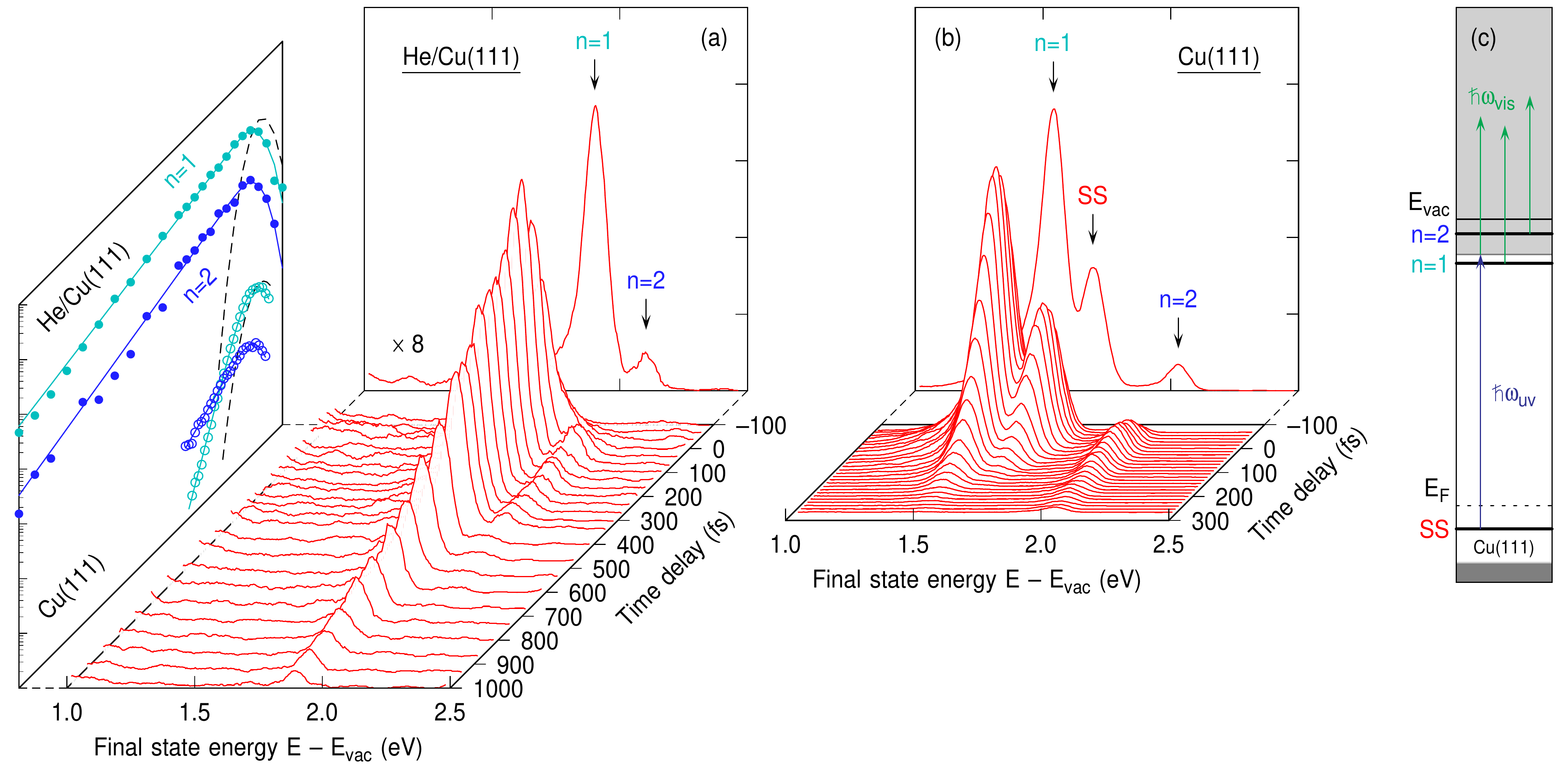}
    \caption[]{2PPE spectra from (a) a saturated He monolayer and (b) the
    pristine Cu(111) surface for different time delays
    $\Delta t$ between pump and probe pulses, according to the
    excitation scheme in (c).
     The rear panels in (a) and (b) show 2PPE spectra for $\Delta
    t\approx 0$ with maxima assigned to the image-potential states ($n=1,2$) and the
    Shockley-type surface state (SS).
    The left panel with logarithmic intensity scale shows
    transient 2PPE data of the ($n=1$)-state (green circles/dots
    for pristine Cu(111)/He monolayer) which decay with single
    exponentials (green solid lines).
     The dashed lines in this panel indicate the cross correlation
    of the pump and probe pulses obtained from the signal of the
    non-resonantly emitted Shockley state for the pristine (bottom)
    and He-covered surface (top).}
    \label{fig:fig:2ppe}
    \end{center}
\end{figure*}

Figure~\ref{fig:fig:2ppe} shows two time-resolved series of
2PPE spectra at normal emission obtained from the He covered
(a) and bare (b) Cu(111) surface.
 Both series show three prominent features.
 The peaks denoted as $n=1$ and $n=2$ are assigned to the first
two members of the Rydberg-like series of initially unoccupied
image-potential states which are populated by the uv pump
pulses and subsequently photoemitted by the vis probe pulses,
as illustrated in (c)\cite{Note2}.
 The signals of the ($n=1$)- and ($n=2$)-states appear
for the bare surface at final state energies of $1.55$ and
$2.05$~eV, respectively (rear panel of (b)).
 This corresponds to binding energies with respect to $E_{\rm vac}$
of $E_{n=1}=0.73\pm0.03$~eV and $E_{n=2}=0.24\pm0.03$~eV.
 We note that theses binding energies are slightly smaller
than values reported for Cu(111) bulk samples
\cite{Wolf96prb,Weinelt02jp,Damm09prb,Borisov06prb}.
 The peak at $1.71$~eV denoted by SS results from
direct photoemission from the partially occupied intrinsic
Shockley-type surface state by non-resonant 2PPE with one uv
and one vis photon and is only visible for temporal overlapping
pump and probe pulses.
 At the $\bar{\Gamma}$-point this state has a binding energy
of $E_{SS}=0.39\pm0.05$~eV with respect to the Fermi level.

\begin{figure}[t]
    \begin{center}
    \includegraphics[width = 0.4750\textwidth]{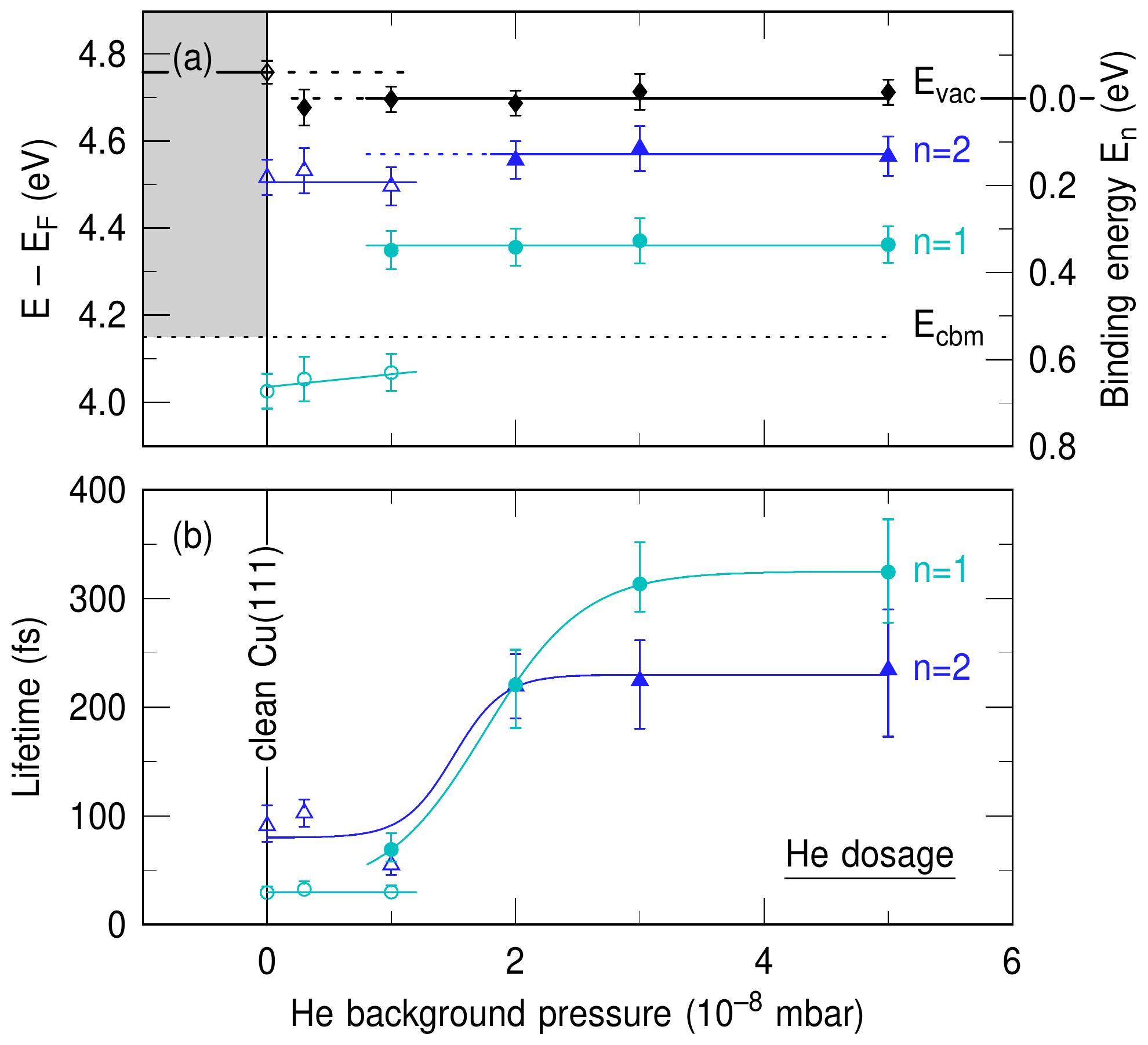}
    \caption[]{(a) Energy with respect to the Fermi level $E_{F}$ and
    (b) lifetimes of the image-potential states $n=1$ and $n=2$ as a function
    of the He background pressure (green/blue data points, the solid lines
    are a guide for the eye). Open (solid) symbols depict data assigned to
    the bare (He covered) surface. Black data points represent the position of
    the vacuum level which serves as reference for the binding energy $E_n$.
    The shaded gray area depicts the projected bulk bands at the $\bar{\Gamma}$-point of Cu(111).}
    \label{fig:fig:dosage}
    \end{center}
\end{figure}

The 2PPE spectra of the He covered Cu(111) surface of
Fig.~\ref{fig:fig:2ppe}(a) have been recorded at a He
background pressure $p_{\rm He}$ of $5\times10^{-8}$~mbar which
supports a coverage of one monolayer under laser irradiation,
as it will be shown below.
 Compared to the pristine Cu(111) surface the signal of the
SS is almost completely quenched and the ($n=1$)- and
($n=2$)-maxima show a considerable blue shift to final states
energies of 1.94 and 2.13~eV with respect to $E_{\rm vac}$,
respectively.
 At these energies, also the ($n=1$)-state is
degenerated with the projected Cu bulk bands
[Fig.~\ref{fig:fig:dosage}] and is in fact an image-potential
resonance.

 The energy shift is almost completely caused by a reduction of the
binding energies by about a factor of two
($E_{n=1}=0.35\pm0.03$~eV, $E_{n=2}=0.16\pm0.03$~eV), whereas
the weakly polarizable He film reduces the work function $\Phi$
only marginally by 50~meV [Fig.~\ref{fig:fig:dosage}].
 Thus, even one monolayer of He strongly decouples the image-potential
states from the metal surface.
 Despite this decoupling, the binding energies are much
larger than those reported for electrons on the surface of LHe
($E_{n=1}\sim 1$~meV \cite{Cole69prl}), because of the strong
electron attraction by the highly polarizable metal substrate
underneath the He film.

 Lifetimes of electrons excited into the image-potential
states at the $\bar{\rm \Gamma}$-point have been determined by
measuring the 2PPE intensity at the respective energies as
function of the time delay between the uv and the vis laser
pulses [Fig.~\ref{fig:fig:2ppe}, left panel].
 The lifetimes have been extracted from best fits (solid
lines) using a rate-equation model assuming single-exponential
population decay.
 On the bare Cu(111) surface, the states $n=1$ and $n=2$
show lifetimes of $\tau_{\rm n=1}=34\pm5$~fs and $\tau_{\rm
n=2}=108\pm15$~fs, respectively \cite{supplemental}.

 Adsorption of a monolayer of He drastically changes the electron dynamics
in the image-potential states:
 $\tau_{n=1}$ increases by one order of magnitude to
$330\pm60$~fs [Fig.~\ref{fig:fig:dosage}(b)] despite the fact
that the $n=1$ state becomes an image-potential resonance.
 The lifetime of the $(n=2)$-state, which is already an
image-potential resonance on the bare Cu(111) surface, is less
affected.
 It increases only by a factor of two to $256\pm60$~fs.

For coverage dependent data, the He density within the
illuminated spot has been calibrated by systematic 2PPE
measurements as a function of $p_{\rm He}$ while keeping the
laser intensities constant.
 The results are summarized in Fig.~\ref{fig:fig:dosage} where
$E_{n=1,2}$ (a) and $\tau_{n=1,2}$ (b) are plotted as a
function of $p_{\rm He}$.
 For increasing $p_{\rm He}$, $\Phi$ slightly drops by only 50~meV.
The peaks assigned to the ($n=1,2$) states on the pristine Cu
surface decrease continuously and vanish for $p_{\rm
He}>1\times10^{-8}$~mbar.
 Simultaneously, two new peaks with lower binding energies appear in the 2PPE
spectrum, which we assign to the first two image-potential
states on the He covered areas.
 Just at $p_{\rm He}=1\times 10^{-8}$~mbar, we still observe the
 ($n=1$)-peak assigned to the uncovered areas and already
 the ($n=1$)-peak assigned to the covered areas \cite{supplemental}.
 This might be surprising at first glance, since submonolayers
of physisorbed He do not grow as 2D islands like the heavier
rare-gases, but form instead a 2D gas.
 In our experiment, the coexistence of covered and uncovered
areas results from the competition of adsorption and laser
induced desorption which causes a variation of the coverage
across the spot profile of the laser.
 The desorption rate is, however, small. We estimate it to be
$<10^{-7}$~ML per laser shot \cite{supplemental}.
 The assignment of the states is consistent with their lifetimes
as a function of $p_{\rm He}$.
 The lifetimes of the states assigned to bare Cu(111) areas
remain almost constant while those of the states assigned to
the He covered areas increase with $p_{\rm He}$ and saturate
above $p_{\rm He}= 2\times 10^{-8}$~mbar, which we interpret as
the formation of a saturated monolayer.
 We do not expect the formation of the more weakly bound second monolayer \cite{supplemental}.

%
 On the Cu(111) surface, the change of the decay dynamics upon
He adsorption is an interplay of decoupling of the wave
functions from the metal and energy shift with respect to the
Cu bulk bands. Whereas the decoupling increases the lifetime,
the shift into resonance with the Cu bulk bands reduces it,
because it opens the additional decay channel of elastic
resonant charge transfer into the bulk
\cite{Borisov06prb,Hofer16ss}.

 The decoupling is caused by the strong Pauli repulsion by the
closed-shell He atoms.
 This repulsion constitutes a high tunneling barrier for the
whole series of image-potential states and extend up to 1.3~eV
above the vacuum level for a thick layer of liquid He
\cite{Sommer64prl}.
 The image force attraction induced by the metal substrate
reduces this barrier, whereas the larger density of the
monolayer due to the much stronger He-metal compared to He-He
dispersion forces increases it \cite{Broomall76prb}.
 The actual barrier height can be estimated from the
above-threshold maximum of the kinetic energy distributions of
secondary electrons, which in good approximation corresponds to
the top of the barrier with respect to the vacuum level.
 Our data and those of Ref.~\cite{Kossler11dr} yield 0.8~eV for
this quantity; in combination with the ($n=1,2$)-binding
energies this results in effective barrier heights of 1.15 and
0.96~eV for the ($n=1$) and ($n=2$)-states, respectively.

 These barriers push the image-potential states further away
from the metal and strongly reduce the wave function overlap
with the Cu bulk and surface states.
 This overlap is the crucial quantity for the lifetime of
image-potential states and resonances, because
 it determines the efficiency of both inelastic and elastic decay
\cite{Echeni04ssr,Borisov06prb}.
 The latter dominates the decay of image-potential
resonances and is responsible for their general much shorter
lifetimes as compared to image-potential states at the same
barrier height \cite{Damm09prb}.
 Against this background, the observed increase of the
($n=1$)-lifetime on Cu(111) by one order of magnitude is
unprecedentedly large and clearly opens up new possibilities
for detailed investigations of 2D electron gases.

 In the framework of the present study, the choice of
Cu(111) films on Ru(0001) as a substrate was dictated by
technical constrains. In a further optimized experimental setup
one would make use of substrates like Cu(100) or Ag(100), on
which the image-potential states are located far from the
projected bulk bands \cite{Chulkov99ss2}.
 The present data allow us to predict the degree of decoupling
by a He layer on these substrates by comparing the measured
($n=1$)-resonance lifetime of He/Cu(111) with that of a
hypothetical resonance on bare Cu(111) at the same energy.
 The dominant elastic contribution of its lifetime can
be calculated reliably within a multiple scattering approach
that provides the electron reflectivity $r_{\rm C}$ at the
surface barrier as a function of energy \cite{Hofer16ss}.
 Potential parameters of Cu(111) \cite{Chulkov99ss2} yield
for this quantity 0.74 (0.65) at an energy of 0.1~eV (0.2~eV)
above the band edge, which corresponds to an elastic lifetime
of only 4~fs (2.9~fs).
 Even if we allow an uncertainty of the exact band edge position in
our Cu(111) films of 0.1~eV, this estimation demonstrates that
the decoupling of a single ML of He is capable to enhance the
($n=1$)-lifetime by up to two orders of magnitude. This is
about 30 times larger as compared to Ar \cite{Berthold02cpl}.

 In conclusion, we have shown that the ultimate model
system of a 2D electron gas on a thin He film grown on an
atomically smooth single crystalline metal surface is
accessible with time-resolved photoemission spectroscopy.
 We find that its lifetime is under such conditions only limited
by tunneling and not by surface defects.
 In a further optimized setup, which reaches temperatures
below 1~K as required for the growth of thicker He layers, our
results let us expect to realize lifetimes of several hundred
picoseconds even on an only 2-ML-thick He film on Cu(100) or
Ag(100).
 The binding energy on such thin film is still large enough to
support high electron densities without becoming instable due
to ripplon formation \cite{Etz84prl}.
 In combination with angle-resolved photoemission spectroscopy,
this opens the possibility to observe phenomena such as Wigner
crystallization and melting into a degenerate 2D electron gas
within momentum space.

We gratefully acknowledge funding by the Deutsche
Forschungsgemeinschaft through Project No.\ GU495/2 and
SFB~1083. SK and PF acknowledge support by the Deutsche
Forschungsgemeinschaft through project No. FE246/2 and by the
Munich-Centre for Advanced Photonics through project No.\ MAP
B.1.4.
%
%
\bibliographystyle{prsty}

\end{document}